\documentclass{article}
\usepackage{charter}
\usepackage{hyperref}

\usepackage[accepted]{sysml2019}
\usepackage{enumitem}
\usepackage{flushend}
\usepackage{amsmath,amsthm,amssymb}
\usepackage{mathtools}
\usepackage[scientific-notation=true]{siunitx}
\usepackage{wrapfig}
\usepackage{xspace}
\usepackage[linesnumbered,lined,ruled,algo2e]{algorithm2e}
\usepackage{hyperref, cleveref, nameref}

\def\xparagraph#1{\vspace{5pt}\noindent\textbf{#1}: }

\usepackage[final]{listings}
\usepackage{multicol}
\usepackage{multirow}
\usepackage{hyperref}
\usepackage{subcaption}
\usepackage[hypcap=true]{caption}
\usepackage{pgfplots}
\usepackage{pgfplotstable}
\usepackage{csvsimple}
\usepackage[all]{xy}
\usepgfplotslibrary{groupplots}
\usetikzlibrary{patterns}
\usepgfplotslibrary{colorbrewer}
\pgfplotsset{compat=newest}
\usepackage{filecontents}

\usepackage[utf8]{inputenc}
\usepackage{lipsum}

\usepackage{soul,color}

\usepackage{xlop}
\usepackage{booktabs}

\usepackage{enumitem}

\sysmltitlerunning{TicTac: Accelerating Distributed Deep Learning with Communication Scheduling}

\begin{document}

\twocolumn[
\sysmltitle{TicTac: Accelerating Distributed Deep Learning with Communication Scheduling}

\sysmlsetsymbol{equal}{*}

\begin{sysmlauthorlist}
\sysmlauthor{Sayed Hadi Hashemi*}{uiuc}
\sysmlauthor{Sangeetha Abdu Jyothi*}{uiuc}
\sysmlauthor{Roy H Campbell}{uiuc}
\end{sysmlauthorlist}

\sysmlaffiliation{uiuc}{University of Illinois at Urbana-Champaign}
\sysmlcorrespondingauthor{Sayed Hadi Hashemi}{hashemi3@illinois.edu}

\sysmlkeywords{Machine Learning, SysML}

\vskip 0.3in
\begin{abstract}
    State-of-the-art deep learning systems rely on iterative distributed training to tackle the increasing complexity of models and input data. The iteration time in these communication-heavy systems depends on the computation time, communication time and the extent of overlap of computation and communication. 

In this work, we identify a shortcoming in systems with graph representation for computation, such as TensorFlow and PyTorch, that result in high variance in iteration time --- random order of received parameters across workers. We develop a system, TicTac, to improve the iteration time by fixing this issue in distributed deep learning with Parameter Servers while guaranteeing near-optimal overlap of communication and computation. TicTac identifies and enforces an order of network transfers which improves the iteration time using prioritization. Our system is implemented over TensorFlow and requires no changes to the model or developer inputs. TicTac improves the throughput by up to $37.7\%$ in inference and $19.2\%$ in training, while also reducing straggler effect by up to $2.3\times$. Our code is publicly available.

 \end{abstract}
]
\printAffiliationsAndNotice{\sysmlEqualContribution} 

\section{Introduction}

Artificial intelligence has grown significantly in the past decade, fuelled by the flexibility of development offered by machine learning frameworks, availability of rich data, and readily accessible distributed high-performance computing. The computational cost of training sophisticated deep learning models has long outgrown the capabilities of a single high-end machine, leading to distributed training being the norm in a typical AI pipeline. Training a deep learning model is an iterative job which may take days to weeks in high-end clusters today.

Computational graphs are used to represent the training jobs in state-of-the-art systems~\cite{abadi2016tensorflow,chen2015mxnet,paszke2017pytorch}. In the commonly-used Model Replica or data parallel mode of training, the input data is partitioned and processed at participating workers using identical computational graphs. Each iteration typically lasts milliseconds to seconds. At the end of each iteration, servers exchange a relatively large amount of data associated with parameter updates to aggregate the results of the iteration. This communication overhead has a substantial impact on throughput of the system and also limits its scalability~\cite{sridharan2018scale,alistarh2017qsgd}. Even a small improvement in communication overhead can improve the learning time by hours in these long-running learning jobs.

The iteration time in deep learning systems depends on the time taken by (i) computation, (ii) communication and (iii) the overlap between the two. When workers receive the parameters from the parameter server at the beginning of each iteration, all parameters are not used simultaneously; they are consumed based on the dependencies in the underlying DAG. While one particular schedule of parameter transfers (over the complete set of parameters in a given model in a single iteration) may facilitate faster computation, another may cause blockage. Hence, identifying the best schedule of parameter transfers is critical for reducing the blocking on computation (determined by DAG dependencies), and in turn improving the overlap and the iteration time.

We observe that the schedule of data transfers in current systems~\cite{abadi2016tensorflow,chen2015mxnet,paszke2017pytorch} is determined arbitrarily during execution without considering the impact on overlap. We quantify the observed combinations in TensorFlow and find that in a trial with $1000$ iterations on ResNet-V2-50, every iteration had a unique order of received parameters which has not been observed previously. This random order of parameter transfers at workers has two performance implications. First, the iteration time, and in turn throughput (number of samples processed per second), suffers significantly due to sub-optimal overlap. Second, even in the same iteration, multiple workers might follow different schedules of data transfers, leading to stragglers during synchronized training.

Past work has attempted to address this issue by enforcing the same order of parameter transfers at all workers. However, these work were restricted to earlier systems with layer-by-layer model representation~\cite{arnold2016ddl,cui2016geeps,203269} where finding the optimal order of execution is trivial~\cite{cui2014exploiting}. In modern systems with DAG representation~\cite{abadi2016tensorflow, paszke2017pytorch}, this is a non-trivial challenge.

In this work, we devise a systematic methodology for deriving near-optimal schedules of parameter transfers through critical path analysis on the underlying computational graph. This allows maximal overlap of computation and communication and prevents stragglers arising from random order of parameter transfers at workers. We also develop a lightweight resource-level enforcement mechanism over TensorFlow~\cite{abadi2016tensorflow}. These techniques form the core of our system, TicTac, which achieves substantial performance improvement while requiring no changes in the model or developer inputs.

In this work, we make the following contributions:
\begin{itemize}[leftmargin=*]
    \itemsep0em
    \item We identify an opportunity for improving performance in state-of-the-art deep learning systems with Parameter Server-based aggregation through finer-grained resource-aware scheduling that solves the problem of random parameter transfers (\S\ref{sec:motivation}).
    \item We define a metric to quantify the scheduling efficiency of a given execution (\S\ref{sec:formal}). 
    \item We propose two heuristics, TIC and TAC, for near-optimal scheduling of computation and communication in Model Replica with Parameter Server. 
    \item We implement our system over TensorFlow (\S~\ref{sec:design}). The code is publicly available.
    \item We extensively evaluate the performance of our system in GPU and high-end CPU environments under training and inference of DNN models. We show that throughput can be improved by up to $37.7\%$ with realistic workloads (\S\ref{sec:results}).
\end{itemize}

 \section{Background and Motivation}
\label{sec:motivation}

\begin{figure}
  \centering
  \includegraphics[width=0.4\columnwidth]{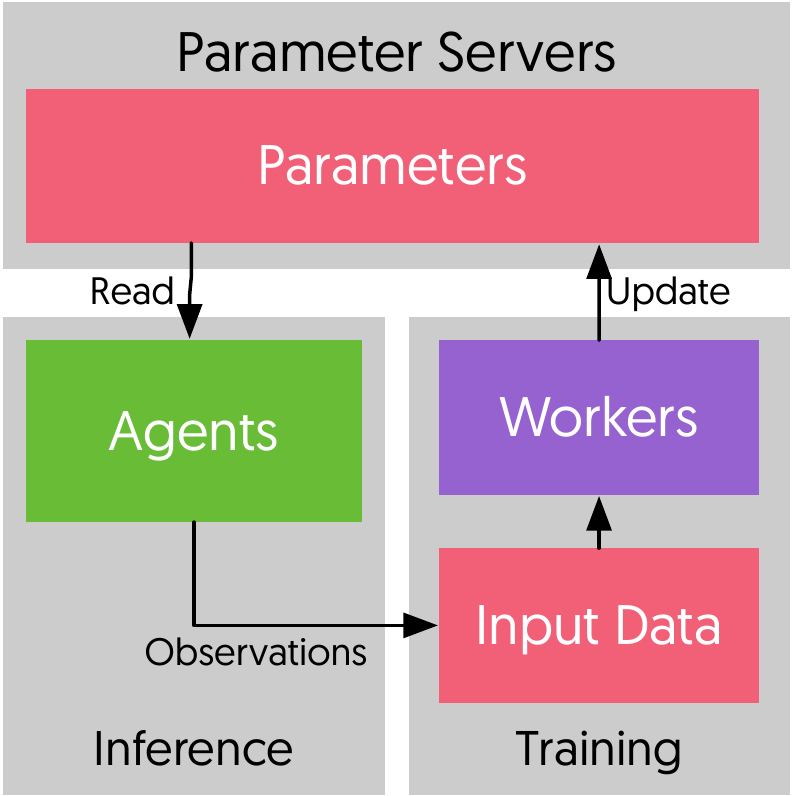}
  \vspace{-2mm}\caption{\small A general reinforcement learning setup }\vspace{-5mm}
  \label{fig:rl}
\end{figure}

Our system focuses on network optimization in deep learning frameworks with DAG representation of computational graphs~\cite{abadi2016tensorflow, paszke2017pytorch}, Model Replica (MR) mode of distribution and Parameter Servers. The performance improvement provided by TicTac is beneficial in two key environments. First, it improves throughput and iteration time in clud environment with commodity hardware or on-demand clusters where high resiliency is critical (workers may be preempted). Second, in online reinforcement learning with workers for training and separate active agents for inference, enforced ordering can improve the inference time. In this environment, the active agents are reading parameters from the PS or decentralized workers as shown in Figure~\ref{fig:rl}. While decentralized aggregation techniques (such as all-reduce and Horovod~\cite{horovod}) are gaining traction in high performance networking, TicTac does not address such systems and is focused on PS.

In this section, we give a brief overview of deep learning systems, prior techniques proposed in these systems to mitigate network overhead, and opportunities for further optimization.

\subsection{Network Optimization in DNN training}
Communication cost, critical in deep learning systems, increases with scale out \cite{sridharan2018scale,alistarh2017qsgd}. In this scenario, for high GPU utilization, it is beneficial to have communication time less than or equal to the computation time. Moreover, efficient overlap of communication and computation is also critical for high throughput. Several techniques have been proposed to improve the system performance.

\xparagraph{Increasing computation time} The fraction of computation time relative to communication time can be increased by increasing the batch size~\cite{iandola2016firecaffe}. However, this approach suffers from decreased accuracy~\cite{keskar2016large} necessitating additional correction mechanisms and may not be generally applicable under resource constraints. \cite{goyal2017accurate,cho2017powerai,DBLP:journals/corr/abs-1709-05011, DBLP:journals/corr/abs-1711-04325}. 

\xparagraph{Decreasing communication time} Solutions for reducing network communication have taken multiple approaches --- modifying the machine learning algorithm to reduce communication cost~\cite{alistarh2017qsgd, wen2017terngrad, 203269}, reducing the precision of parameter representation~\cite{vanhoucke2011improving,courbariaux2015binaryconnect,gupta2015deep}, changing the network primitives to collective (e.g. all reduce)~\cite{goyal2017accurate, cho2017powerai,  deepspeech2, DBLP:journals/corr/abs-1709-05011, DBLP:journals/corr/abs-1711-04325} or broadcast \cite{203269}. 

\xparagraph{Smarter interleaving of computation and communication} This approach is adopted by several layer-by-layer systems \cite{arnold2016ddl,cui2016geeps,203269} where the models are sequential and obtaining the order is trivial \cite{cui2014exploiting}. These solutions are not generally applicable to current DAG-based systems such as TensorFlow~\cite{abadi2016tensorflow} and PyTorch~\cite{paszke2017pytorch}. The inter-resource dependency considered in \cite{cui2016geeps} (with GPU memory) and in \cite{203269} (with network) is constrained to layer-by-layer models. 

In this work, we focus on \textit{improving the iteration time through better and predictable overlap of communication and computation in Parameter Server.} Techniques for optimizing communication and communication time are orthogonal to our system and may be used in parallel with TicTac.

\subsection{Opportunity for Optimization}
In MR, each worker has an identical copy of the computational DAG. In addition to the worker DAG, we also have a DAG at the PS which is different from that at the worker.  \textbf{PS DAG} has five ops per parameter: aggregation, $send$, $recv$, read, and update. Aggregation, read and update on PS are typically lightweight. The transfers are driven by the workers as PS activates all $send$ and $recv$ at the beginning of each iteration. Since the load on PS is dominated by network transfers, the problem of communication-computation overlap does not arise in PS. In the \textbf{worker DAG}, all \texttt{recv} ops are roots and $send$ ops are leaves. Thus $recv$ ops can block the initialization of a computation branch in the DAG. Since the activation of various branches of computation in the DAG is dependent on the $recv$ at the root of the branch, the ordering in MR can be reduced to the ordering of \texttt{recv} ops in workers.

\begin{figure}
    \centering
    \begin{tabular}[t]{cc}
\begin{subfigure}[b]{0.35\columnwidth}
    \centering
    \includegraphics[width=\textwidth]{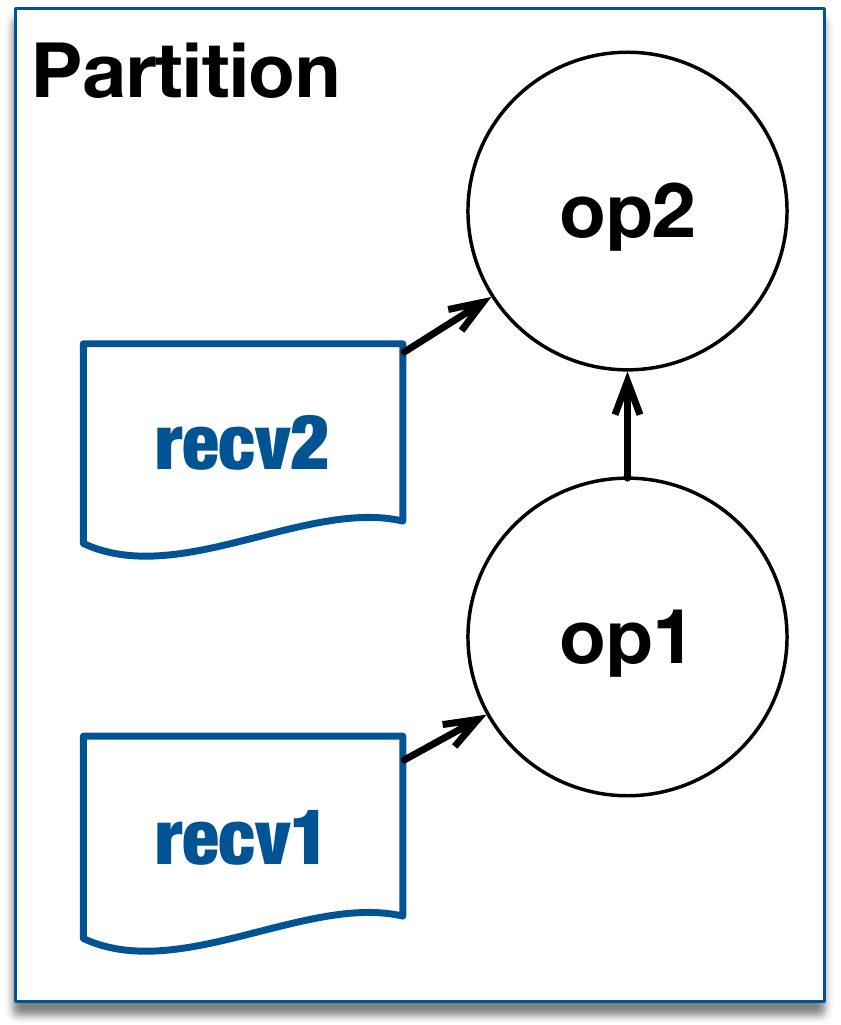}
    \caption{\small Toy Computational Graph} 
    \label{fig:example-dag}
\end{subfigure}
        \begin{tabular}[b]{@{}c@{}}
        \smallskip
            \begin{subfigure}[t]{0.57\columnwidth}
                \centering
               \includegraphics[width=\textwidth]{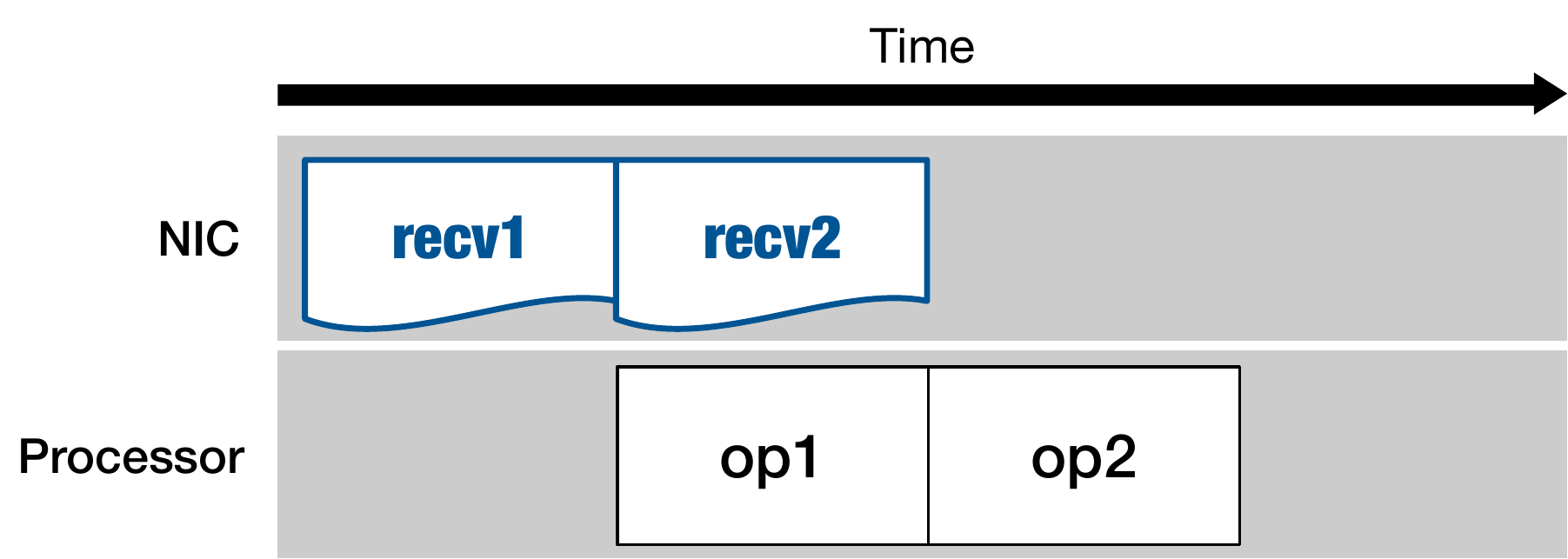}
                \caption{\small Good Execution Order}
                \label{fig:example-best}
            \end{subfigure}
            \\
            \begin{subfigure}[t]{0.57\columnwidth}
                \centering
                \includegraphics[width=\textwidth]{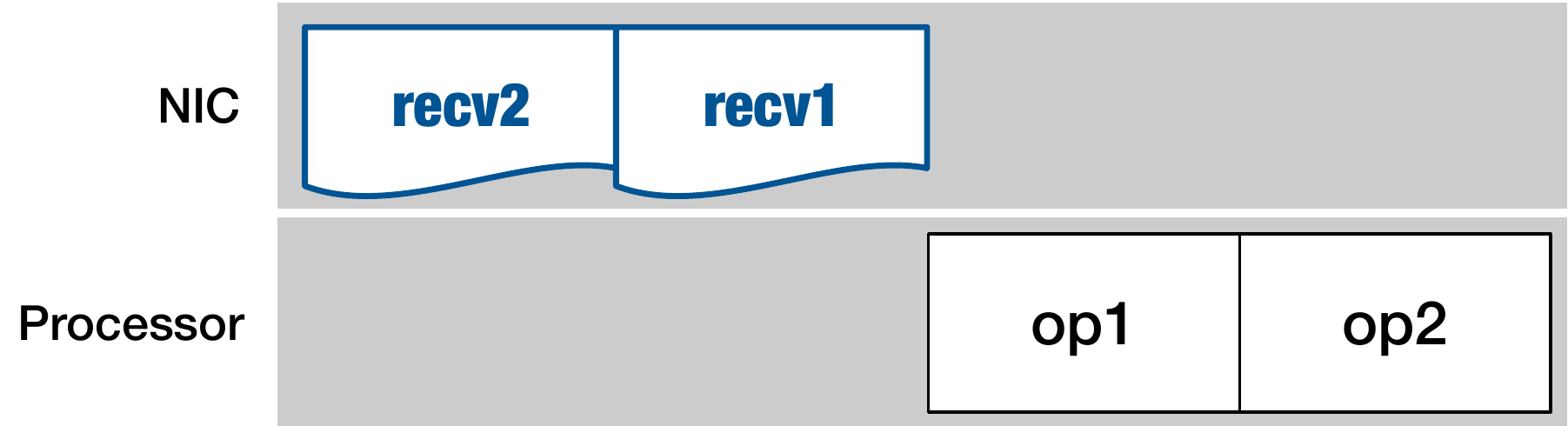}
                \caption{\small Bad Execution Order}
                \label{fig:example-worst}
            \end{subfigure}
        \end{tabular}
    \end{tabular}
    \vspace{-3mm}\caption{\small Impact of multi-resource operation ordering on performance}\vspace{-5mm}
    \label{fig:example}
\end{figure} For example, in the simple DAG shown in Figure~\ref{fig:example-dag}, there are two possible schedules for parameter transfers. If $recv_1$ is transferred before $recv_2$, it reduces the blocking on computation time and improves the overlap. The reverse order results in increased iteration time due to blocking on computation.

Thus, in a distributed environment, network can block computation depending on dependencies in the DAG. This can lead to under-utilization of computational capacity, in turn resulting in sub-optimal performance. In addition, variation in iteration time across multiple devices can lead to straggling effect. 

The impact of poor overlap can be significant in DNN training due to complexity of state-of-the-art models. For instance, ResNet-v2-152~\cite{DBLP:journals/corr/HeZR016} has $363$ parameters with an aggregate size of $229.5$MB. The computational graph associated with this neural network has $4655$ operations in the TensorFlow framework. Finding the optimal schedule in this complex DAG involves evaluating 363! combinations. We run $1000$ iterations of learning over ResNet-v2-50, Inception-v3 and VGG-16 networks and observe the order of network transfers at a single worker. The observed order of parameter transfer is unique in ResNet-v2-50 and Inception-v3 networks across the $1000$ runs. In VGG-16, we observe $493$ unique combinations across $1000$ runs.

\subsection{Comparison with Other Distributed Systems}
It is worth noting that deep learning systems with computational graphs are fundamentally different from graph processing systems \cite{malewicz2010pregel,hoque2013lfgraph,xin2013graphx}. In deep learning, the graph is a representation of the computation to be done on the input data. In graph processing systems, the graph itself is the input to the computation. As a result, graphs in DNN frameworks are a few orders of magnitude smaller than a typical large-scale graph processing system. Iterations in DNN frameworks are identical, and network communication pattern is fixed. This may not be true for graph processing systems. 

In stream processing systems, the relationship between processing elements are represented using graphs. These systems allow pipelining, with different partitions of input data being processed in different elements along the pipeline at the same time. In contrast, DNN frameworks process the entire batch of input at a processing element at a worker. Pipelining is not employed in this environment. Hence, optimizations proposed for stream processing cannot be borrowed here.

\begin{figure}
  \centering
  \includegraphics[width=\columnwidth]{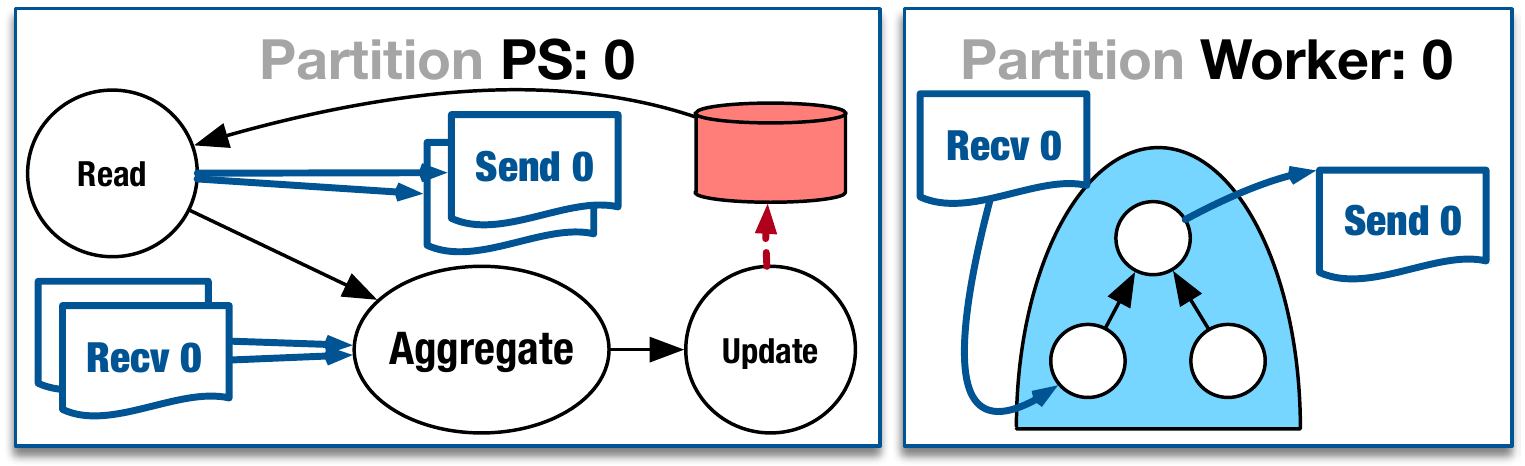}
  \vspace{-6mm}\caption{\small Distributed execution of Model-Replica with Parameter Server}\vspace{-5mm}
  \label{fig:mr}
\end{figure}

 \section{Scheduling Efficiency}
\label{sec:formal}
In this section, we explore methods for quantitatively comparing the efficiency of multiple schedules. Towards this goal, we formally define the scheduling problem and investigate the feasibility of finding an optimal solution. Finally, we define a metric that is used to quantify the efficiency of a schedule.

\subsection{Scheduling Problem}
The objective is to find the optimal schedule of network transfers that minimizes the iteration time by improving the communication/computation overlap. The network transfers ($recv$ ops) are roots in the computational graph. The branch of computation ops dependent on a $recv$ op can be executed only after the network transfer is completed. Thus, the order of network transfers can determine the order of computation as well as the extent of overlap. We focus on improving the overlap, and in turn the iteration time, by choosing a near-optimal schedule of ops.

\textbf{The inputs} to this optimization problem are: (a) \textit{the partitioned graph}, and (b) the characteristics of the underlying platform represented by \textit{a time oracle}. The partitioned graph is the computational graph with resource tags associated to each op --- computation ops are assigned to computation resources, communication ops to corresponding communication channels. The time oracle~($Time(op)$) predicts the execution time of a given op. For computation ops, this indicates the elapsed time on a  computation resource. For communication ops, this represents the transfer time on the communication medium. We compute the time assuming that the resource is dedicated to the op under consideration.

\textbf{The output} of the scheduling algorithm is a feasible schedule of ops that minimizes the iteration time. Recall that ops in a computational DAG may have multiple feasible topological orders. However, some of them may result in a bad iteration time (as explained in Figure~\ref{fig:example}). We want to limit the execution path to take only a subset of this valid set of orders that improves the training performance. We achieve this with priority numbers. Priority number is a positive integer assigned to an op in the partitioned graph. A higher priority op is given a lower priority number. An op may not be assigned a priority if it need not be ordered. Multiple ops may be assigned the same priority if their relative order is insignificant.

During distributed training, the order is enforced in the following manner. When a resource needs to select a new item from the ready-to-execute queue, it randomly chooses from among the set of ops that contain the lowest priority number and those without any priority number. It is worth noting that priority only specifies relative order among candidate ops in the ready-to-execute queue at a given resource, and the resulting order will still respect the topological order specified by the computational DAG.

The problem of finding the optimal schedule in an MR model can be mapped to finding the minimum makespan in a job shop problem with dependencies (flowshop in \cite{doi:10.1287/moor.1.2.117}), where machines can be mapped to resources, tasks to ops and task dependencies to the computational DAG dependencies.
The solution to this problem is known to be NP-Hard \cite{doi:10.1287/moor.1.2.117}, therefore we propose an approximate solution as a heuristic algorithm in \S\ref{sec:solution}. Note that we focus on prioritizing ops on individual resources within a single device, given an assignment of ops on devices. The widely-studied multi-resource/multi-processor scheduling is orthogonal to our problem and deals with assigning tasks to individual components, but does not consider relative priorities between them \footnote{It should be noted that assigning devices using multi-resource scheduling is still an open problem in distributed machine learning system. It is beyond the scope of this paper.}.

\subsection{Scheduling Efficiency Metric}
\label{sec:boundry}
We define a metric, Scheduling Efficiency ($E(G,Time,makespan)$), to measure the effect of scheduling and variation in makespan from other sources. Makespan is the total time taken by the schedule (in one iteration). This metric is used to evaluate the quality of our heuristics since we do not have a solution to the NP-hard scheduling problem.

To define this more formally, we introduce upper and lower boundaries of makespan with respect to the schedule. Note that these bounds ignore the dependencies in the DAG, and hence may not be achievable in practice. The inputs are the computational graph, $G$, set of resources, $D$, and the time oracle, $Time$, which gives the execution time of each operation, $op$, as $Time(op)$.

The worst makespan (the longest) is computed by assuming only one resource is utilized at any given moment during the execution, i.e., the ops are executed sequentially. The upper bound on makespan is given by:
\begin{equation}
	U_{Makespan}(G, Time)=\sum_{\mathclap{op \in G}}Time(op)
\end{equation}
Note that in practice, the achieved makespan will be lower than this value since computation and communication can happen in parallel. 

The lower bound of the makespan is computed by assuming all the resources are always utilized (without any restrictions imposed by the DAG). The lower bound on makespan is given by:
\begin{equation}
	L_{Makespan}(G, Time)=\max_{d\in D}\sum_{op \in G_d}Time(op)
\end{equation}
where $G_d$ refer to all ops assigned to the resource $d$. Note that this may not be achievable in practice since an op assigned to a resource may have to wait for its dependencies to complete before it can execute. Even if the target resource is available, the op's dependencies may be executing on a different resource.

For a given iteration, we measure runtime of each op ($Time(op)$) as well as the makespan of that iteration (\textit{m)}) and then calculate the bounds on makespan. We define Scheduling Efficiency as follows:
{\smaller
\begin{equation}
	E(G,Time, m) = \frac{U_{Makespan}(G,Time) - m}{U_{Makespan}(G,Time) - L_{Makespan}(G,Time)}
\end{equation}}
$E=1$ indicates a perfect ordering, and $E=0$ means the worst ordering.

Next, we define \textit{Speedup}, \textit{S(G,Time)}, as the maximum theoretical performance speedup possible with the best schedule relative to the worst schedule:
{\smaller
\begin{equation}
	S(G,Time) = \frac{U_{Makespan}(G,Time) - L_{Makespan}(G,Time)}{L_{Makespan}(G,Time)}
\end{equation}
}
This metric quantifies the benefits achievable on a given DAG with an efficient schedule. $S=0$ indicates no benefit from scheduling, and $S=1$ means double the throughput. Generally, if one resource on a device has significantly higher load than others, then better scheduling will have less effect on the makespan since we are restricted by the bottleneck resource. In practice, when one resource is significantly takes longer than the other resource, the scheduling benefits are limited.

 \section{Scheduling Algorithms}
\label{sec:solution}

In this section, we present two heuristics to approximate the optimal schedule of $recv$ ops in a reference worker partition (\S\ref{sec:formal}). The intuition in our heuristics is to prioritize transfers that speed up the critical path in the DAG. The heuristics are:

\xparagraph{Timing-Independent Communication scheduling (TIC)} In TIC, we assign priorities based only on vertex dependencies in the DAG (ignoring the execution time of each op). Higher priorities are given to transfers which are least blocking on computation. In this algorithm, we ignore the time oracle, $Time$, and assume all ops have equal cost. 

\xparagraph{Timing-Aware Communication scheduling (TAC)} In this algorithm, we prioritize transfers that maximize the computation/communication overlap by using information on (a) execution time of each op estimated with time oracle and (b) dependencies among ops specified by the computational DAG.

\subsection{Op properties} 
Before delving into the algorithms, we define properties associated with ops that are used in the scheduling algorithms. The algorithms are given a partitioned graph ($G$), a time oracle ($Time$), available communication channels on a device ($C$) and a set of outstanding (to-be-activated) $recv$s ops ($R$). We assume that $recv$ ops not in $R$ have their corresponding transfers completed. These properties are updated using the algorithm~\ref{alg:properties}.

\xparagraph{Communication Dependency ($\mathbf{op.dep}$)} We define \texttt{communication dependency} of an op as the set of $recv$ ops that the op is directly or transitively dependent on ($op.dep$). For example, in figure~\ref{fig:example-dag}, $op_2.dep = \{ recv_1, recv_2 \}$. We extract the communication dependencies using a depth-first post-fix graph traversal on the DAG.

\begin{algorithm2e}[ht]
\smaller
  \SetAlgoLined
  \SetKwProg{Fun}{Function}{}{end}
  \tcp{Update properties for the given the set of outstanding read ops \textit{R}}
  \Fun{UpdateProperties($G$, $Time$, $R$):}
  {
  \ForEach{$op \in G$}
  {
  $op.M\leftarrow \sum_{\forall r \in op.dep \cap R} Time(r)$\;
  }
  \ForEach{$op \in R$}
  {
  $op.P\leftarrow 0$\;
  $op.M^+\leftarrow +\infty$\;
  }
  \ForEach{$op \in G-R$}
  {
  $D\leftarrow\textit{op.dep}\cap R$\;
  \If{$|D| = 1$}{
  $\forall r \in D: \textit{r.P} \leftarrow \textit{r.P} + \textit{Time(op)}$\;
  }
  \If{$|D| > 1$}{
    $\forall r \in D: r.M^+ \leftarrow \min \{r.M^+, op.M\}$\;
  }
  }
  }
e  \caption{Property Update Algorithm}
  \label{alg:properties}
\end{algorithm2e}
 
\xparagraph{Com\underline{m}unication Time ($\mathbf{Op.M}$)} Communication time of an op is the total network transfer time required to complete that op. For a $recv$ op, this is the time required to complete its corresponding transfer, given by $Time(recvOp)$. For other ops, this is the total time to complete all outstanding dependent transfers, given by \\ $ \sum_{ r \in op.dep \cap R } Time(r)$. For example, in Figure~\ref{fig:example-dag}, $op_1.M = Time(recv_1)$ and $op_2.M= Time(recv_1) + Time(recv_2)$.

\vspace{2mm}
\noindent For \textbf{$\mathbf{recv}$ ops}, we define two additional properties.

\xparagraph{Directly-Dependent Com\underline{p}ute Load ($\mathbf{recvOp.P}$)} This property represents the computational benefit of completing a $recv$ op. More specifically, it is the total $Time(op)$ for all ops that can be activated only by completing this $recvOp$, but not without it. These ops are those whose communication dependencies contain only this outstanding $recvOp$ (it is admissible to have communication dependencies on other completed $recv$ operations). For example, in Figure~\ref{fig:example-dag}, $recv_1.P = Time(op_1)$ and $recv_2.P = 0$ since no op can execute with completion of only $recv_2$.

\xparagraph{Impending Communication Load ($\mathbf{recvOp.M^+}$)} This property helps us to identify candidate $recv$ ops to be activated, given the current $recv$ is completed. In more detail, it is the minimum communication cost to activate a computation op which has multiple $recv$ dependencies including the one under consideration. For example, in Figure~\ref{fig:example-dag}, $read_1.M^+ = read_2.M^+ = Cost(read_1) + Cost(read_2)$. Please note that $recvOp.M^+$ includes the communication time of that recvOp.

\subsection{Timing-Independent Communication Scheduling (TIC)}

The goal of this algorithm is to prioritize those transfers which reduces blocking on network transfers. Our intuition is that information on DAG structure alone can provide significant improvement.

To achieve this goal, we define a generic time function which only uses the number of communication ops instead of time taken by an op. We use this simple cost function to generate the schedule in Timing-Independent Communication scheduling (TIC).

\xparagraph{General Time Oracle} We define a simple universal time oracle as follows:
{\small
\begin{equation}
  \textit{Time}_\textit{General}(op) = 
  \begin{cases}
    0  & \quad \text{if } op \text{ is not recv}\\
    1  & \quad \text{if } op \text{ is recv}
  \end{cases}
\end{equation}
}

The complete solution is given in Algorithm~\ref{alg:simple}.

\begin{algorithm2e}
\smaller
  \SetAlgoLined
  \SetKwProg{Fun}{Function}{}{end}

\Fun{TIC($G$)}{
  \textit{FindDependencies(G)} \;
  \textit{UpdateProperties(G,R,Time=\{Computation: 0, Communication: 1\})}\;
  $\forall{\textit{op in G, if op is recv}}:  op.priority \leftarrow op.M^+$\;
  }  
\caption{\smaller Timing-Independent Communication Scheduling (TIC)}
  \label{alg:simple}
\end{algorithm2e}

\subsection{Timing-Aware Communication Scheduling (TAC)} 
The goal of this algorithm is to prioritize those transfers which reduces the blocking of computation, i.e., speeding up transfers on the critical path. To achieve this goal, the algorithm focuses on two cases. First, it considers the opportunity for overlapping communication and computation. Second, in the case of equal overlap or absence of it, it looks at the impending transfers to choose one which eliminates the computation block sooner.

To better describe the logic, we begin with an example for each case.  

\begin{figure}
\centering
\begin{subfigure}[t]{0.20\textwidth}
\centering
  \includegraphics[width=\textwidth]{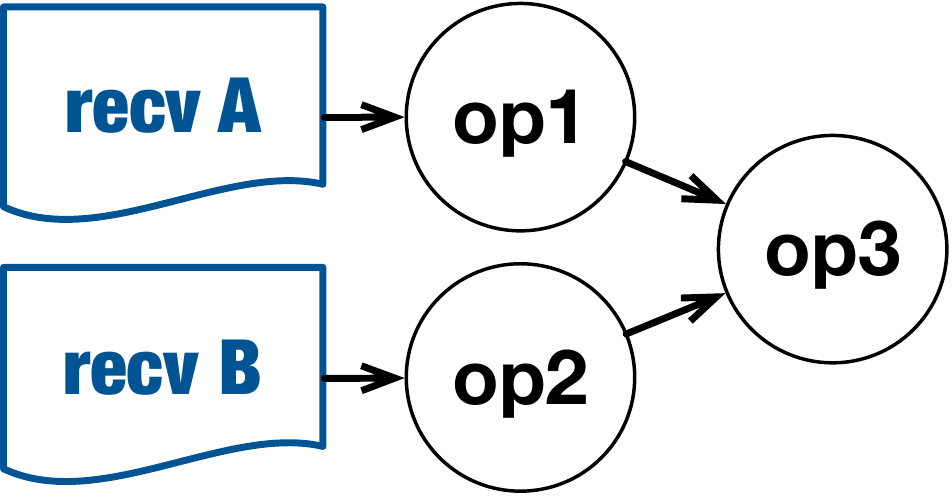}
  \caption{\small Case 1}
  \label{fig:example-dag1}
\end{subfigure}
\begin{subfigure}[t]{0.20\textwidth}
\centering
  \centerline{\includegraphics[width=\textwidth]{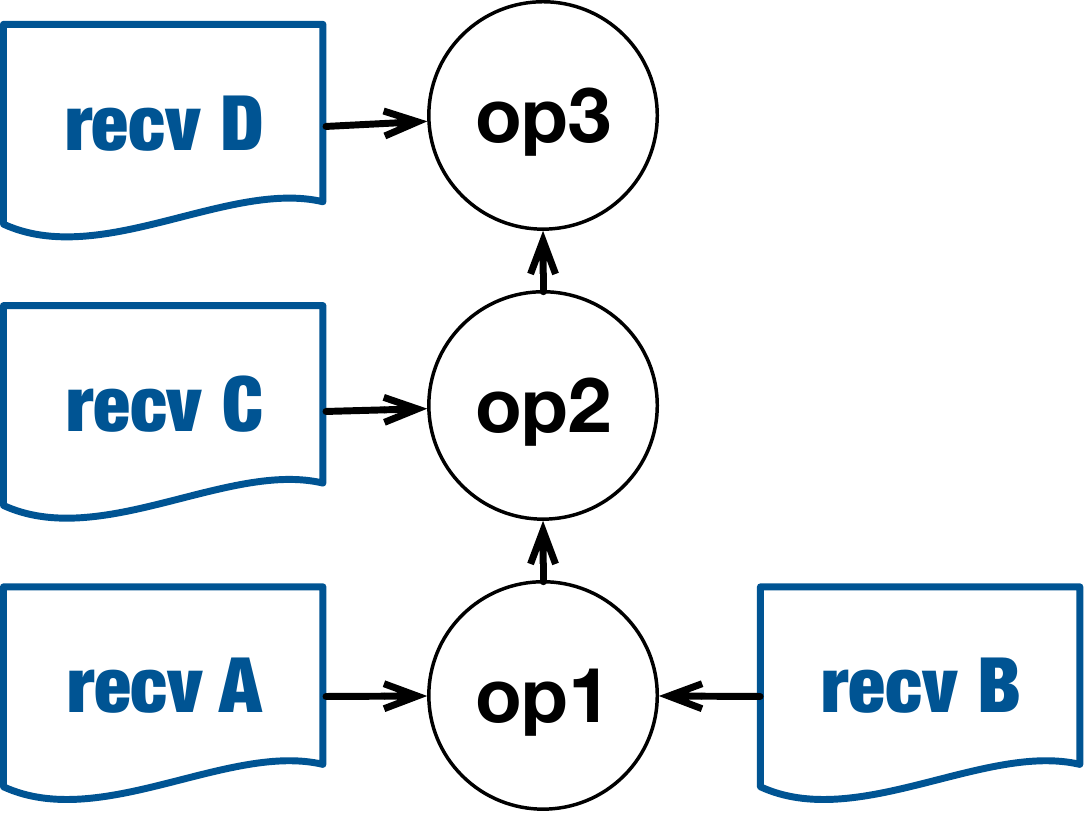}}
  \caption{\small Case 2}
  \label{fig:example-dag2}
\end{subfigure}
  \vspace{-3mm}\caption{\small Sample DAG}\vspace{-5mm}
  \label{fig:example-dag3}
\end{figure}

\xparagraph{Case 1} In Figure~\ref{fig:example-dag1}, when deciding between two read ops, $A$ and $B$, $A$ should precede $B$ iff:
{\smaller
\begin{align*}
 & \mathbf{A \prec B}  \iff Makespan(A\rightarrow B) < Makespan(B \rightarrow A) \\ 
  & \iff M_A + \max\{P_A, M_B\} + P_B < M_B + \max\{P_B, M_A\} + P_A \\
  & \iff M_A + P_A + M_B - \min\{P_A, M_B\} + P_B < \\ & \quad  \quad  \quad \quad M_B + P_B + M_A - \min\{P_B, M_A\} + P_A \\
  & \iff \min\{P_B, M_A\} < \min\{P_A, M_B\}
\end{align*}
}
Therefore:
\begin{equation}
\label{eq:cmp}
  A \prec B \rightarrow \min\{P_B, M_A\} < \min\{P_A, M_B\}
\end{equation}

\xparagraph{Case 2} In Figure~\ref{fig:example-dag2}, when all $recv$ ops are outstanding, their $P$ is $0$, making them equivalent under the comparison in Equation~\ref{eq:cmp}. Obviously, $recv_A$ and $recv_B$ should precede other $recv$s.  Hence, we use $M^+$ to break the ties:  $recv_A.M^+ = recv_B.M^+ = Time(recv_A) + Time(recv_B) < recv_C.M^+ < recv_D.M^+$.  
 
\xparagraph{Comparator} We combine results from the two cases to make a comparator that extends to multiple read ops. This is an approximate induction, which may not be correct in general.
The result is the \texttt{Comparator} function in algorithm~\ref{alg:full}. It is easy to prove that this function is transitive and can be used for partial ordering.

The ordering algorithm takes a partition graph on a worker, calculates the communication dependencies, then while there is an outstanding $recv$ op, it updates properties, finds the smallest $recv$ op with respect to the comparator and then removes the $recv$ from the outstanding set and assign it a higher priority relative to others.

\begin{algorithm2e}
\smaller
  \SetAlgoLined
  \SetKwProg{Fun}{Function}{}{end}
  \tcp{Compare two given $recv$ ops}
  \Fun{Comparator($Op_A$,$Op_B$): Bool}{
  $A\leftarrow\min(P_A,M_B)$\;
  $B\leftarrow\min(P_B,M_A)$\;
  \uIf{$A\neq B$}{
  	\Return $A<B$
  }
  \Else{
  	\Return $M^+_A < M^+_B$
  }
  }
  
\Fun{TAC($G$,$Time$)}{
  \textit{FindDependencies(G)} \;
  $R \leftarrow \{op | \forall \textit{op in G, op is recv}\}$\; 
  $count \leftarrow 0$\;
  \While{\textit{R} is not empty}{
    \textit{UpdateProperties(G,R,Time)}\;
    Find the minimum $op$ from $R$ wrt \textit{Comparator}\;
  Remove $op$ from $R$\;
  $op.priority \leftarrow count$\;
  $count \leftarrow count + 1$\;
  }
}
\caption{\smaller  Timing-Aware Communication Scheduling (TAC)}
\label{alg:full}
\end{algorithm2e}

 \section{System Design}
\label{sec:design}

In this section, we provide a brief overview of the system design and implementation. 

\begin{figure}
    \centering
    \includegraphics[width=0.6\columnwidth]{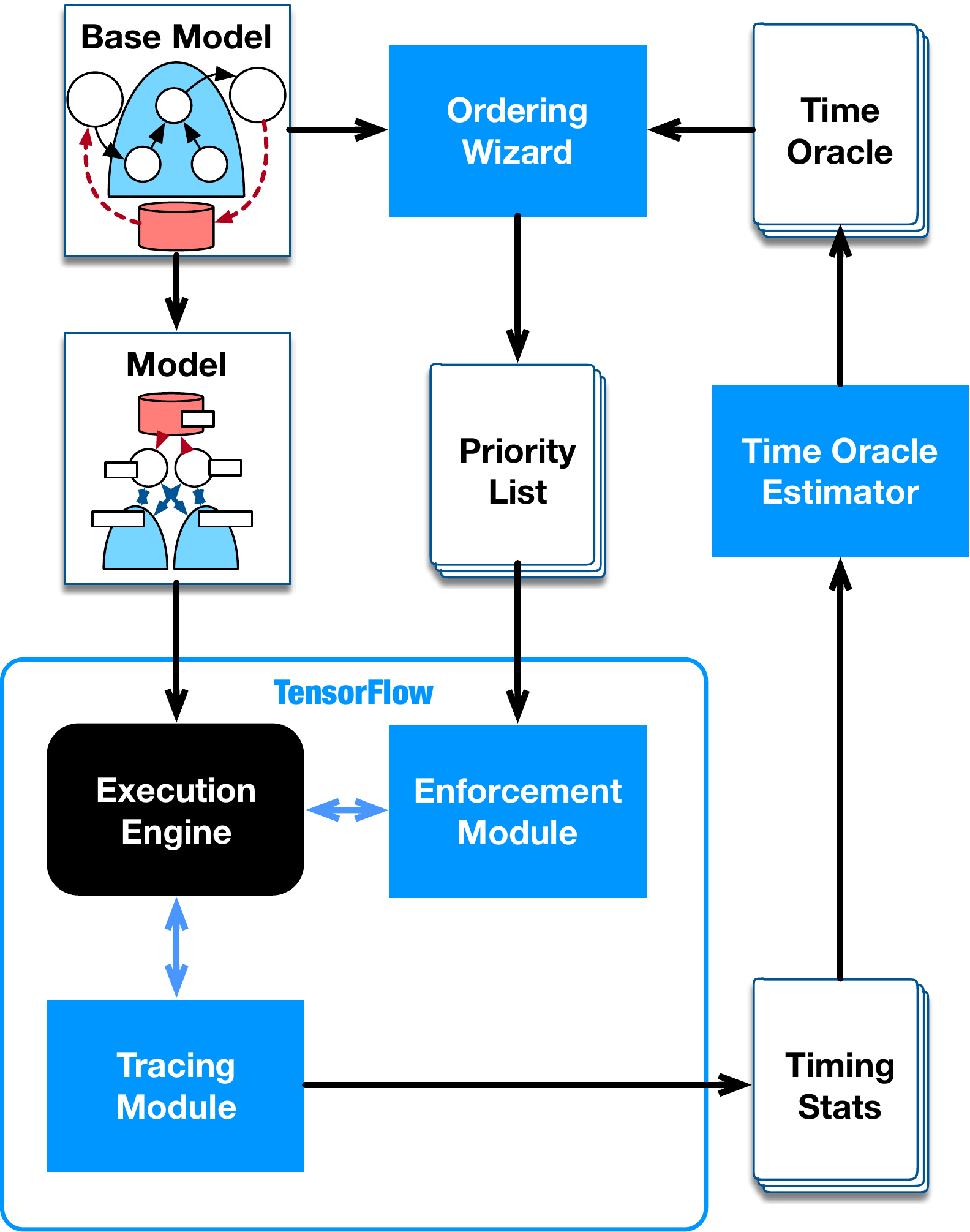}
    \vspace{-3mm}\caption{\small \centering System Design. Components of our system are in blue sharp-edged rectangles. }\vspace{-3mm}
    \label{fig:arc}
\end{figure}

The system has four main components: the tracing module, the time oracle estimator, the ordering wizard, and the enforcement module (shown in Figure~\ref{fig:arc}).

\xparagraph{Tracing Module} This module collects runtime stats from an execution, which is later fed to the time oracle estimator.

\xparagraph{Time Oracle} The time oracle is responsible for estimating the runtime of each op in the system based on the execution timing stats. Note that the runtime may vary depending on the platform, device characteristics, input data and even across iterations on the same hardware/software. We execute each operation $5$ times and measure the time taken in each run. Our Time Oracle implementation chooses the minimum of all measured runs for a given op as the time for that op. 

\xparagraph{Ordering Wizard} This module is responsible for assigning priorities to recv ops on a single worker. The schedule may be computed based on TIC or TAC. In TAC, the ordering module relies on the time estimated by the time oracle. In TIC, the order is determined based on the DAG alone. The estimated priorities are sent to the enforcement module. The priority list is calculated offline before the execution; all iterations follow the same order. 

\xparagraph{Enforcement Module} This module takes as input the priority list computed by the ordering module and enforces this order on the network transfers per worker. 

\subsection{Implementation}
\label{sec:implementation}
We implement our system over TensorFlow 1.8. We describe our implementation in detail.

\xparagraph{Time Oracle}
We use the TensorFlow internal tracer to measure the time of computation ops. We extend the capability (115 LOC C++) of this tracer to collect information on network transfer at all workers. Our code is publicly available (obfuscated for review).

\xparagraph{Ordering Wizard}
We implement TIC and TAC as offline analyzers (250 LOC in Python). 
The implementation takes time oracle and base model in the TensorFlow DAG format and generates the priority of recv ops.

\xparagraph{Enforcing} The enforcement module is implemented over the gRPC submodule of TensorFlow (40LOC in C++).
\begin{figure}
  \centering
  \includegraphics[width=0.7\columnwidth]{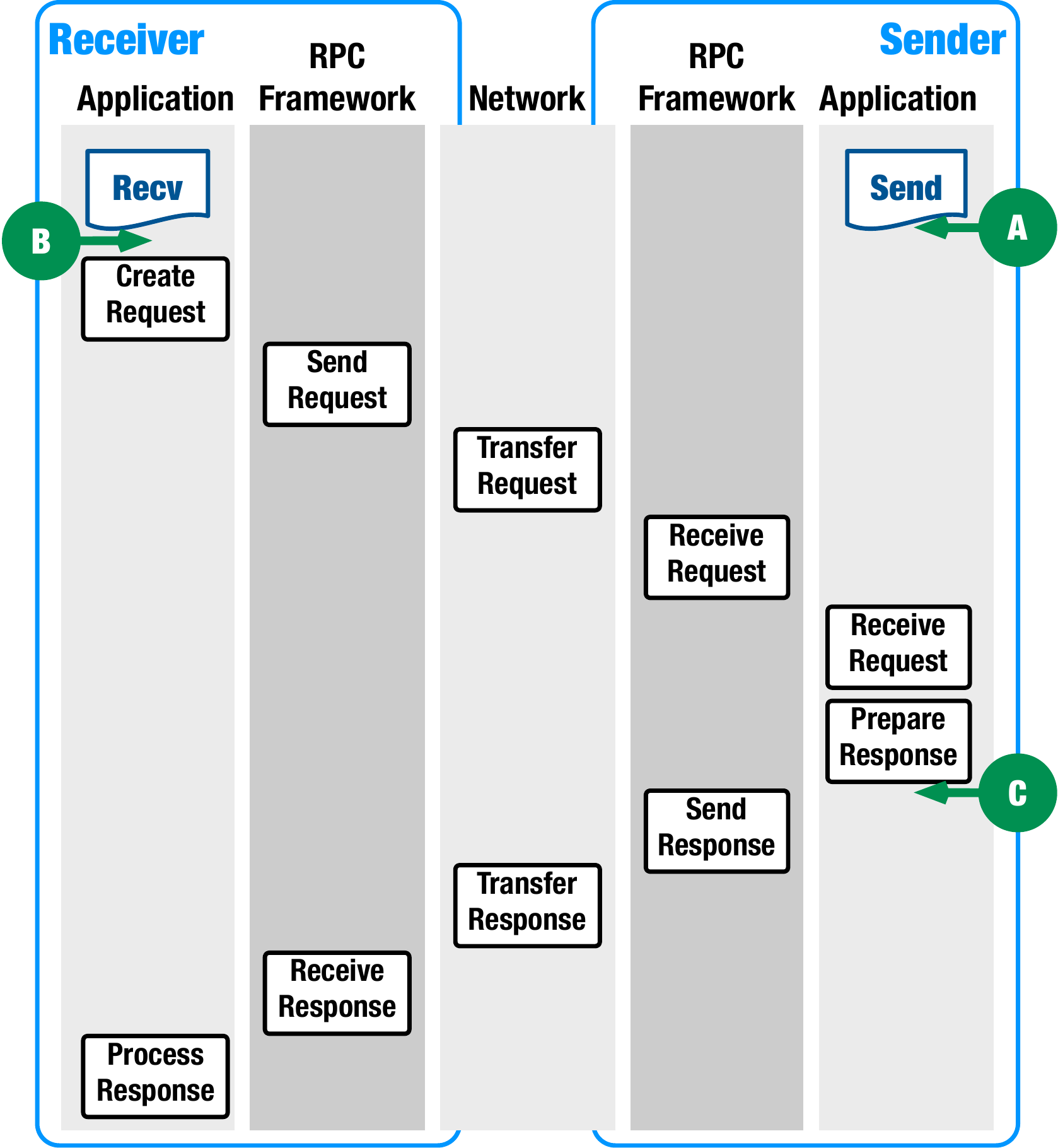}
  \vspace{-3mm}\caption{\small Life time of a network transfer.}\vspace{-3mm}
  \label{fig:rpc}
\end{figure}

gRPC provides one channel per worker-PS pair with all transfers between the pair sent to the same queue. Only one transfer can be active at a given moment for each channel. A network transfer over gRPC in TensorFlow involves multiple stages as shown in Figure~\ref{fig:rpc}. When a recv op is activated at the receiver, it sends a request for transfer to the sender. If the send op is also active at the sender, the transfer may be initiated by gRPC. In this dataflow, there are three possible candidate locations for enforcing ordering --- at the receiver before the request is initiated, at the sender before the send op is activated or at the sender before the transfer is sent to gRPC. Alternatively, this may also be enforced as a direct dependency in the DAG.

We implement the enforcement module at the sender before the transfer is sent to gRPC. This choice is guided by several practical concerns. Enforcing directly on the DAG is conservative since each transfer has to wait for the completion of the previous transfer. This prevents pipelining and drastically reduces the communication throughput. Ordering the activation of recv or send ops is not sufficient since it could change throughout the data flow. For example, a larger transfer request may take longer to reach the response state on the sender side. During this interval, a smaller transfer with lower priority may catch up.

For the purpose of enforcement, the priorities are sequentially assigned to an integer in the range of $[0,n)$. Thus, the priority number of a transfer represents the number of transfers that have to complete before it. The sender (PS server) maintains a counter for each worker per iteration which is incremented when a corresponding transfer is handed to the gRPC. Before a transfer is handed to the gRPC, it is blocked until the corresponding counter reaches the normalized priority number. 

During experiments, we notice that gRPC may not always process transfers in the order they are queued. This affects the performance of our ordering in some cases. However, the number of such occurrences at the gRPC level are very few. In Inception model (one of the tested models), this error was $0.5\%$ in TIC and $0.4\%$ in TAC.

 \section{Results}
\label{sec:results}
In this section, we evaluate TicTac under a wide range of inputs/system parameters to answer the following questions:
\begin{itemize}[leftmargin=*]
\itemsep0em
    \item How does TicTac perform with scale out of workers?
    \item How is TicTac affected by the number of parameter servers?
    \item How does the benefits accrued with TicTac change with the communication and computation cost?
  \item How well do the proposed heuristics perform in terms of consistency and straggler mitigation?
\end{itemize}

\begin{figure*}
    \centering
    \includegraphics[width=\textwidth]{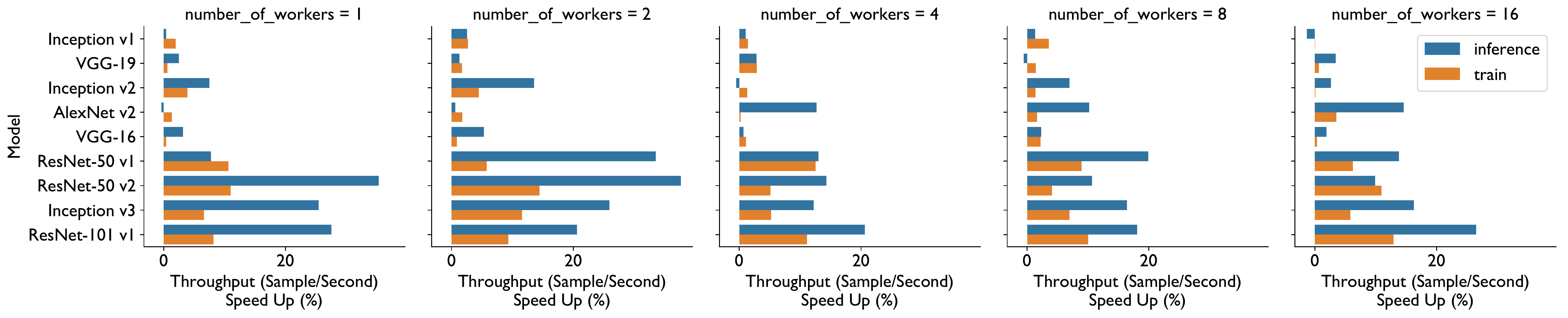}
    \vspace{-8mm}\caption{\small Impact of scaling the number of workers on throughput. The gains are measure with respect to the baseline (no scheduling). Measured on $env_G$ with PS:Workers in the ratio 1:4.}\vspace{-5mm}
    \label{fig:workerScaling}
\end{figure*}

\xparagraph{Setup} We use in-graph replication for Distributed TensorFlow \cite{dist-tensorflow} with synchronized training and synthetic input data. 

We test TicTac under two environments. (a) \textbf{Cloud GPU environment($env_G$):} We use Standard NC6 virtual machines (6 cores, 56 GB memory, 1 X Nvidia K80 GPU with 12GB memory) on Azure cloud environment. For parameter servers we used Standard F64s v2 (CPU Only, 64 cores, 128 GB memory). (b) \textbf{High-end CPU cluster ($env_C$):} We use a commodity cluster (32 core, 64GB memory, 1GbE network). In both environments, we test $2$ to $16$ workers and $1$ to $4$ PS. For understanding the impact of batch size, we test the networks with the standard batch size multiplied by factors [$0.5$, $1$, $2$]. We tested our method on $10$ well-known models (details of models in Table~\ref{tab:models} in Appendix).

We evaluate the performance under two workloads: training and inference. In training, we use Stochastic Gradient Descent (SGD) as optimizer. The training workload is identical to the training jobs used in practice. We emulate the inference workload of agents in reinforcement learning with online training. In this environment, parameter servers store the parameters which are updated by a set of training worker nodes (which we do not consider in the inference workload). The inference agents are responsible for reading the parameters from the PS and running the inference (this is the phase we evaluate in this workload).

In each test, we discard the first $2$ iterations to limit the warm-up effect (initialization of GPUs, cache etc). This is necessary since the first iteration takes much longer compared to the rest. We record the next $10$ iterations. For throughput, we report the mean across $10$ iterations; for straggler effect and scheduling efficiency we report the maximum. Computing the TIC and TAC heuristics takes approximately $10$ seconds. Note that these heuristics are computed \textit{before} the training/inference begins. Hence, this will not add overhead during the execution.

\begin{figure}[h]
\centering
    \includegraphics[width=0.6 \columnwidth]{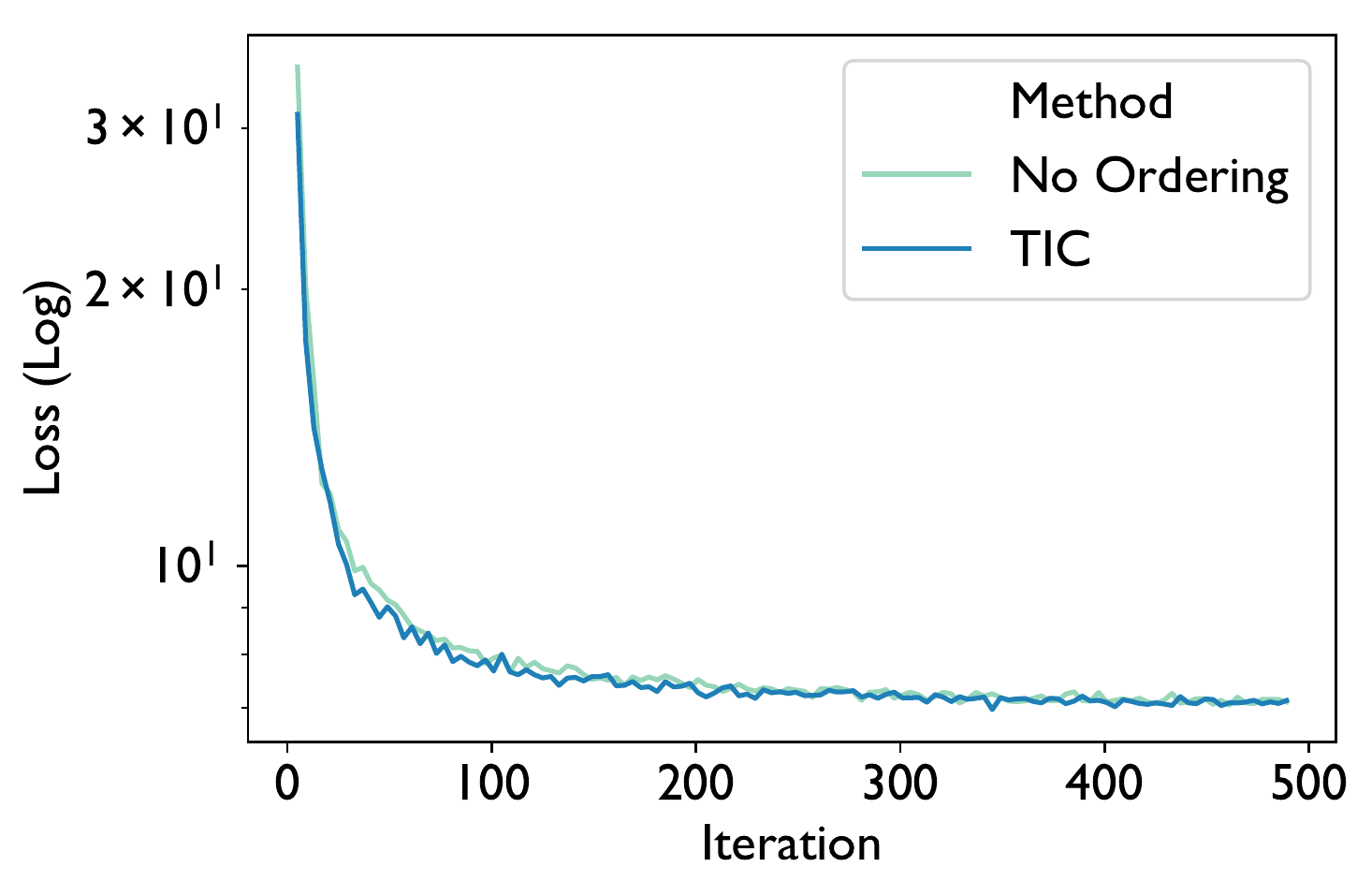}
    \vspace{-1mm}\caption{\small Loss value throughout the first 500 iterations of training InceptionV3 on ImageNet.}\vspace{-2mm}
    \label{fig:loss}
\end{figure}

We use Imagenet Dataset for our experiments. We evaluated both synthetic and real data and observed less than $3\%$ difference in iteration time on a single machine. The data is read in the TFRecord format from a shared NFS-connected Azure storage, samples are resized, augmented, and prefetched during training. TicTac does not alter the computational flow of the model; it only chooses one of the feasible orders of network transfers. Hence, it does not affect the accuracy of training (shown in Figure~\ref{fig:loss}).

Next, we compare the performance metrics across various heuristics. Specifically, we evaluate throughput, scheduling efficiency, and prevalence of stragglers (slow workers that force others to wait, thereby increasing the iteration time). Performance of TIC is only marginally worse compared to TAC (shown in Figure~\ref{fig:cpuRate} in Appendix). This indicates that, for current models, DAG-level information is sufficient for obtaining a near-optimal scheduling. However, we expect the gap between TIC and TAC to increase as complexity of models increases.  

We attempted to compare TicTac with Poseidon~\cite{203269}. However, only the binaries of Poseidon are publicly available. In our experiments, Poseidon performed extremely poorly compared to TicTac, and even vanilla TensorFlow 1.8. Since Poseidon is based on older version of TensorFlow (TensorFlow 1.0) and CUDA (8.0), we were unable to account the poor performance to their methodology. Hence,
we exclude the results since the comparison is inconclusive. Additionally, since order extraction is not explained in their paper, we were unable to reimplement their strategy.

\subsection{Throughput}

\xparagraph{Scaling the number of workers} In Figure~\ref{fig:workerScaling}, we evaluate the impact of scaling the number of workers with the number of PS to workers fixed to the ratio 1:4. We obtain up to $37.7\%$ of speed up in throughput across networks. The gains are measured relative to the baseline --- no scheduling. Larger networks have higher performance gains. The speed up depends on two factors --- communication load and extent of overlap. As the number of workers increases, the communication load increases in PS. When the communication load increases, scheduling can provide benefits through better overlap until a threshold. When the communication load is much higher than the computation load, the impact of overlap diminishes. Hence, beyond this threshold, the benefits accrued with scheduling reduces. This threshold varies across models. Also, since the gains are measured with respect to the baseline which chooses a random schedule, leading to variations in performance. Hence, we observe varying trends across networks based on the network-specific characteristics. In small networks, with small number of workers and parameter servers, the overhead associated with scheduling may overshadow the benefits of better overlap. In such rare cases, we observe a slow down of up to $4.2\%$. This shows that scheduling network transfers may be disabled in small networks at small training and inference sizes.

\begin{figure}
    \centering
    \includegraphics[width=\columnwidth]{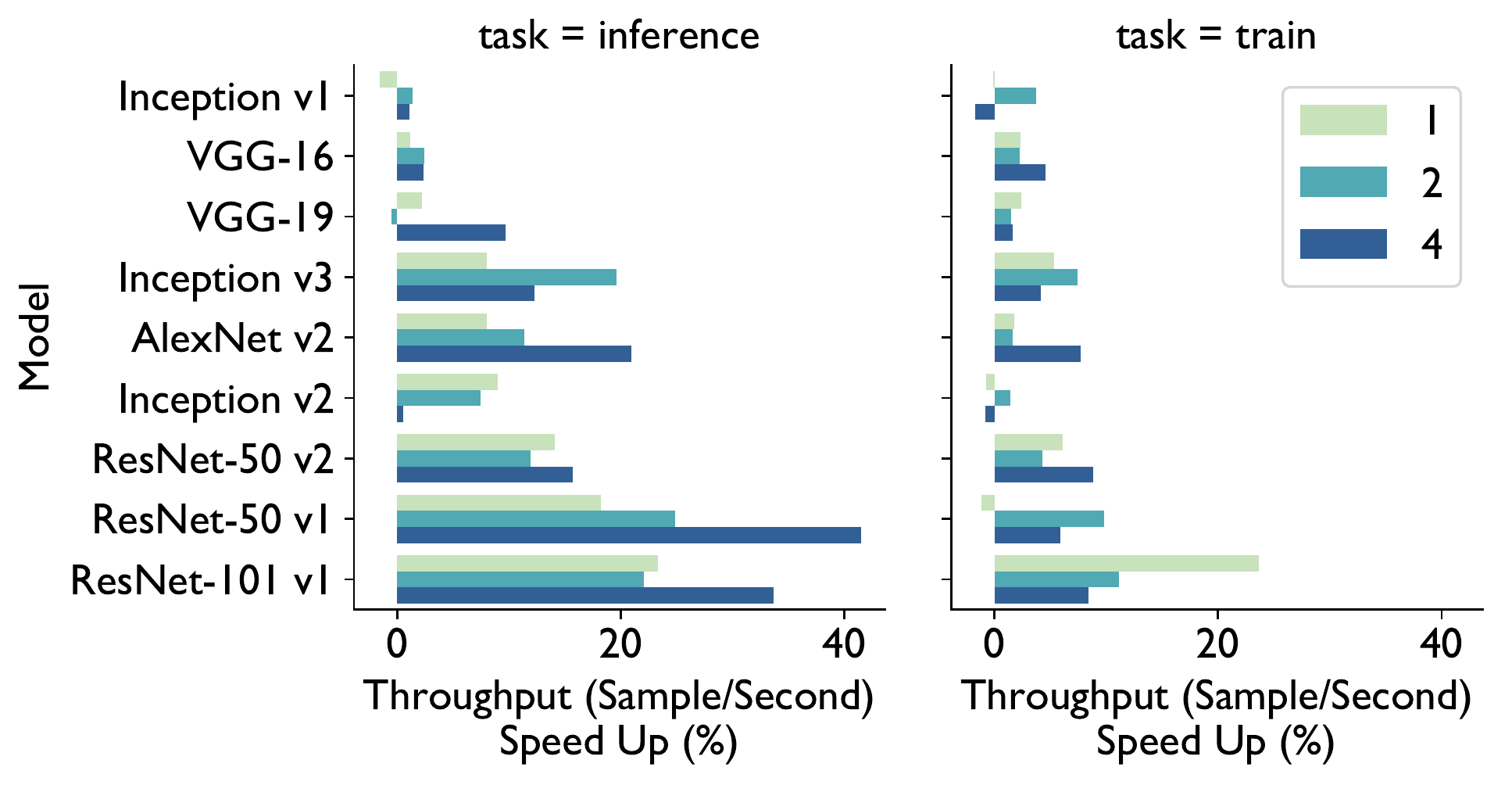}
    \vspace{-8mm}\caption{\small Impact of scaling the number of Parameter Servers on $env_G$ cloud GPU environment with $8$ workers.}\vspace{-5mm}
    \label{fig:psScaling}
\end{figure}

\xparagraph{Scaling the number of Parameter Servers} 
In Figure~\ref{fig:psScaling}, we evaluate the impact of scaling the number of parameter servers with $8$ workers in $env_G$ (Cloud with GPU) across various networks. In general, we obtain higher gains in the inference phase than training. Larger networks obtain higher gains. Even in the presence of multiple parameter servers, enforcing ordering with TicTac provides significant performance improvement.

\xparagraph{Scaling the computational load} 
We study the impact of varying computational load by testing each model with the prescribed batch size multiplied by three factors --- $0.5$, $1$, 
$2$. There are two factors affecting the scaling of computation load --- computation time and opportunity for overlap. The relative ratio of communication and computation determines the opportunity for overlap. As the batch size increases, the computation time increases. If the communication time is higher (compared to the computation time), increase in computation time increases the opportunity for overlap. If communication time is smaller than computation time, scaling will reduce throughput as the opportunity for overlap reduces.

\begin{figure}
\centering
    \includegraphics[width=0.6 \columnwidth]{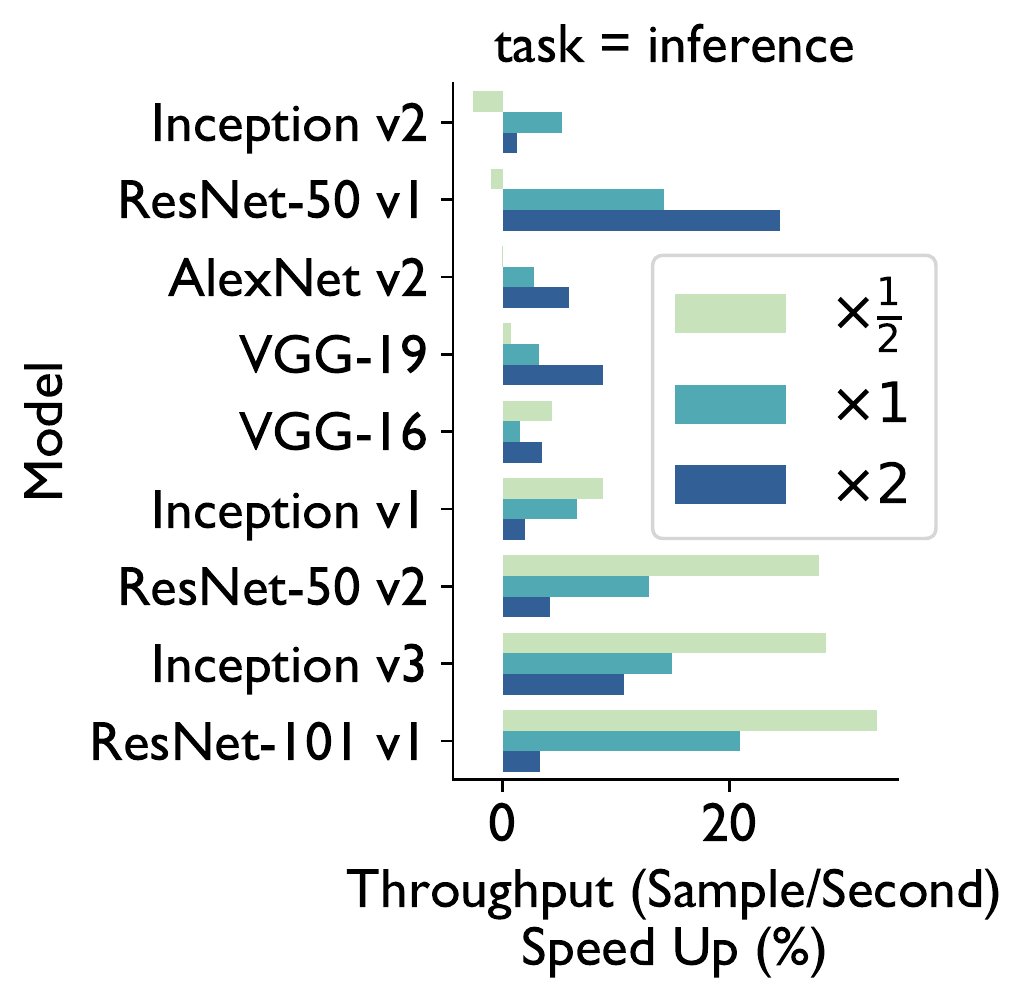}
    \vspace{-3mm}\caption{\small Impact of scaling the computational load on $env_G$ cloud GPU environment with $4$ workers.}\vspace{-3mm}
    \label{fig:batchScaling}
\end{figure}

\subsection{Scheduling Efficiency}

\begin{figure}
\centering
\begin{subfigure}[t]{0.49\columnwidth}
    \centering
    \includegraphics[width=\textwidth]{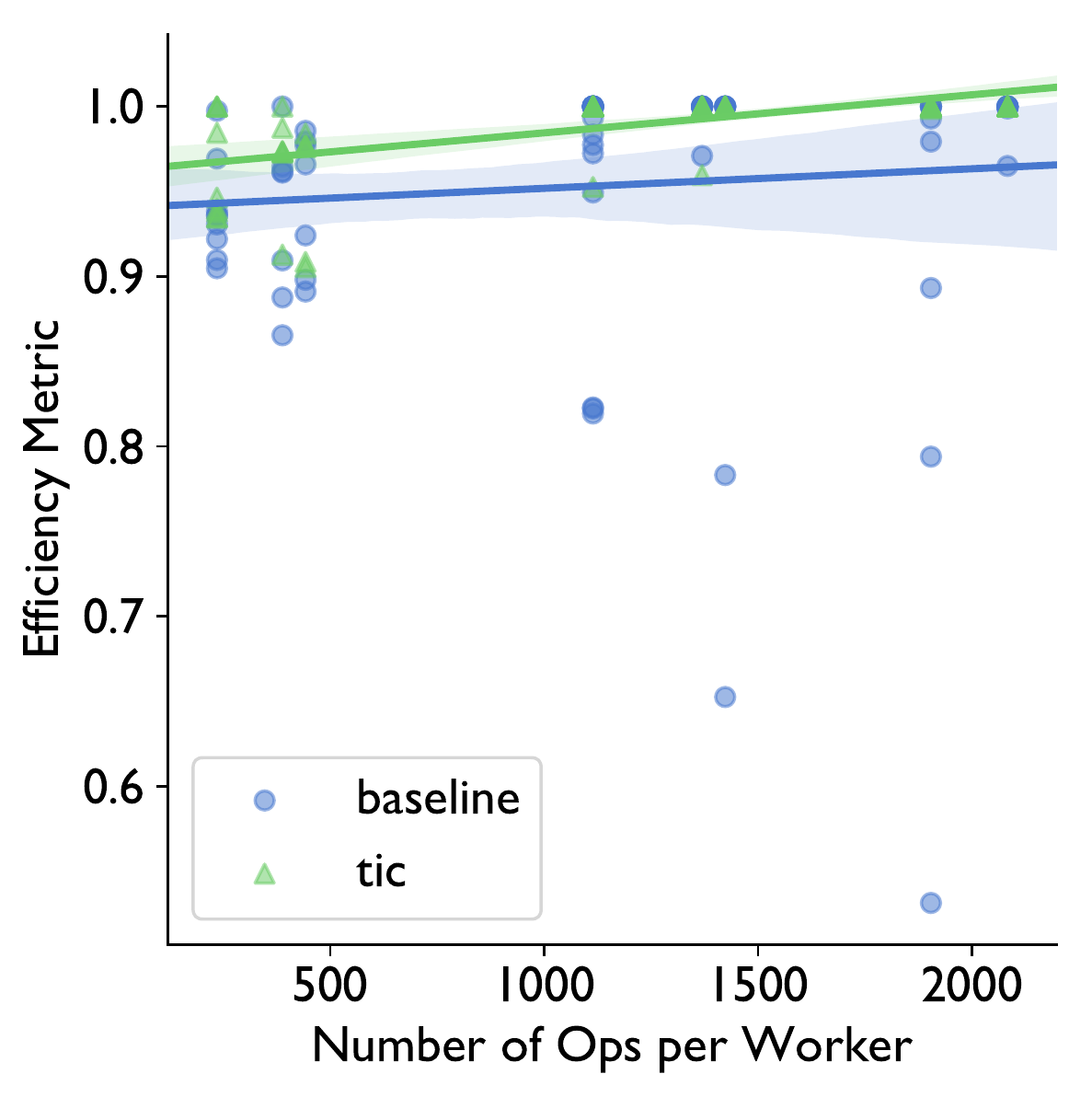}
    \label{fig:eff_g}
\end{subfigure}
\begin{subfigure}[t]{0.49\columnwidth}
    \centering
    \includegraphics[width=\textwidth]{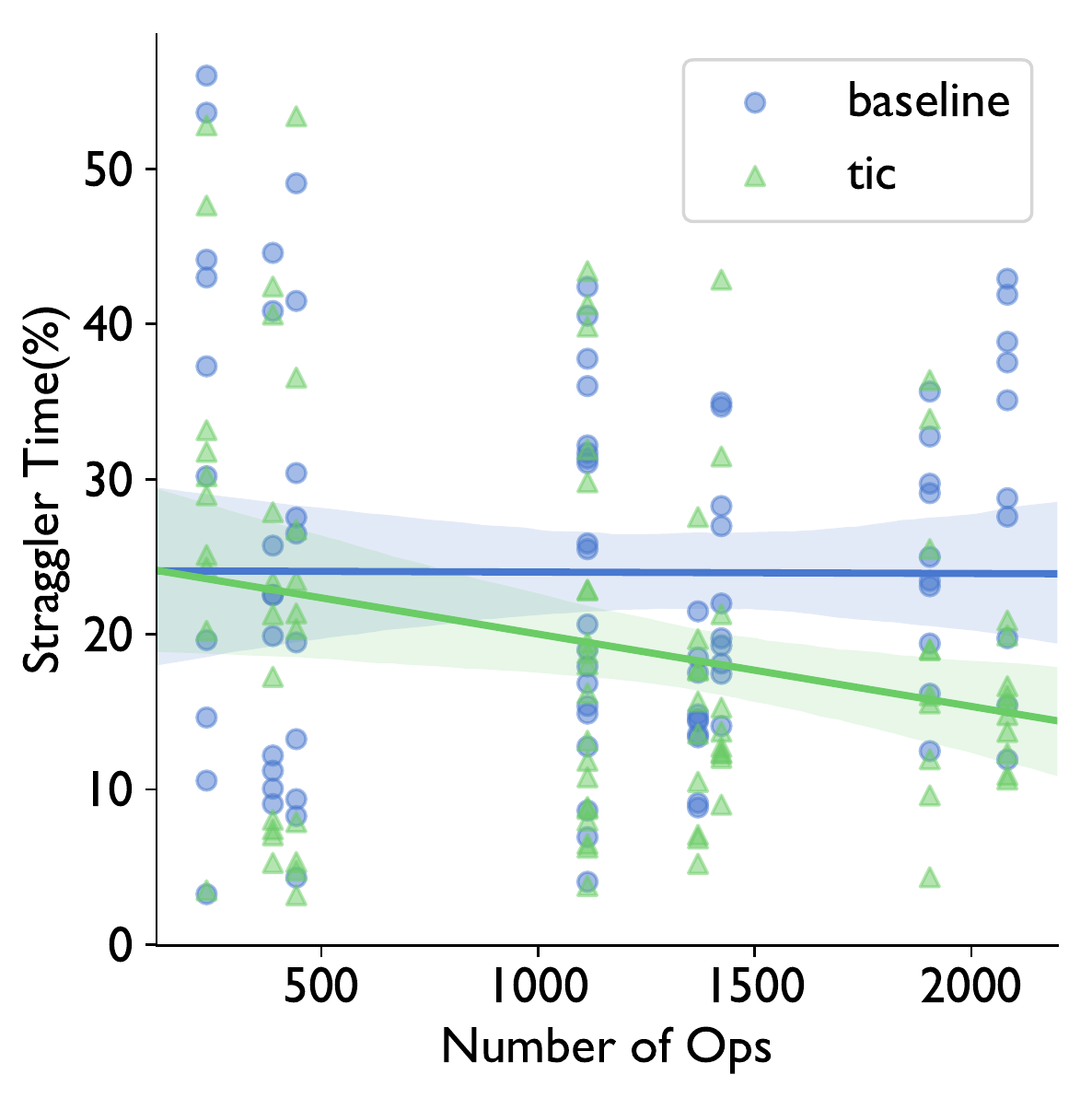}
    \label{fig:straggler}
\end{subfigure}
\vspace{-3mm}\caption{\small On samples including both training and inference in the GPU enviornment, $env_G$, (a) Efficiency metric and  (b) effect of stragglers.}\vspace{-5mm}
\label{fig:compare2}
\end{figure}

To validate the scheduling efficiency metric, we run training of Inception v2 $1000$ times each with and without the scheduling algorithm, TAC in $env_C$. Scheduling efficiency metric can predict step time accurately, with a high $R^2$ score of $0.98$, as seen in Figure~\ref{fig:compare1} (a). This proves that most of the variation in iteration time arises from random schedules in parameter transfers. We also observe that in the absence of enforced scheduling, the step time and scheduling efficiency have a large variance. With scheduling, the step time is reduced and the variance is minimal. Moreover, most runs have a scheduling efficiency approaching $1$, indicating near-optimal scheduling in TAC.

We also measure the scheduling efficiency metric across all models in the cloud environment ($env_G$) with TIC under both training and inference (shown in Figure~\ref{fig:compare2} (a)). We observe that across all models, in all environments, the efficiency metric approaches $1$ even with the simpler timing-independent scheduling mechanism. This is the key factor contributing to the increased throughput.

\subsection{Performance Consistency}

\begin{figure}
\centering
\begin{subfigure}[t]{0.49\columnwidth}
\centering
\begin{tikzpicture}
  \begin{axis}[
    xlabel={Scheduling Efficiency},
    ylabel={Normalized Step Time},
  width=\textwidth,
    legend entries={No Ordering, TAC, LR $(R^2=0.98)$},
    legend cell align=left,
    grid=major,
    ytick={0.5, 0.6,...,1.1},
    legend cell align=left,
    legend style={nodes={scale=0.75, transform shape}},
    legend style={at={(1.25,1.25)},anchor=north},
    legend columns=3, 
    tiny
  ]
  \addplot+[only marks, fill opacity=0.07,draw opacity=0.1, legend image post style={draw opacity=1, fill opacity=0.5}] table[col sep=comma,header=false,x index=0,y index=1]{data/inception-distribution-none.csv};
  \addplot+[only marks, fill opacity=0.07,draw opacity=0.1, legend image post style={draw opacity=1, fill opacity=0.5}] table[col sep=comma,header=false,x index=0,y index=1,on layer=background]{data/inception-distribution-full.csv};
  \addplot+[mark=none, densely dashed, ultra thick, domain=0:1, black, on layer=foreground] {-0.4531309 * x + 0.93852214};
  \end{axis}
\end{tikzpicture}
\label{fig:result-lr}
\end{subfigure}
\begin{subfigure}[t]{0.49\columnwidth}
\centering
\begin{tikzpicture}
\begin{axis}[
  tiny,
  xlabel=Normalized Step Time,
  ylabel=CDF,
  xtick=,
  grid = both,
  legend cell align=left,
  legend style={nodes={scale=0.75, transform shape}},
  width=\textwidth,
]
\def\datafilename{data/inception-cdf.csv}
\addplot+[mark=none, thick] table[col sep=comma,x index={0},y index={1}, header=false] {\datafilename};
\addplot+[mark=none,densely dashed, thick] table[col sep=comma,x index={0},y index={2}, header=false] {\datafilename};
\end{axis}
\end{tikzpicture}
\label{fig:cdf}
\end{subfigure}
\vspace{-3mm}\caption{ \small In $env_C$, on Inception v2, (a) Regression test of Scheduling Efficiency and Normalized Step Time, (b) Step Time Comparison across Scheduling Mechanisms.}\vspace{-5mm}
\label{fig:compare1}
\end{figure}
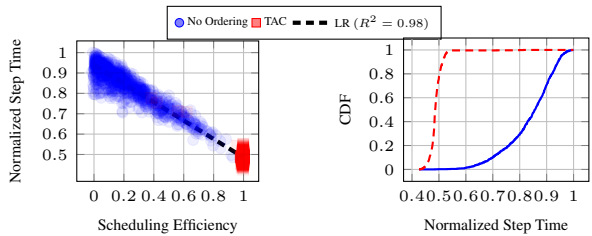
 In Figure~\ref{fig:compare1} (b), we compare the consistency in performance obtained with and without scheduling (TAC) in inference on InceptionV2 with $1000$ runs in $env_C$. We see that both TAC has consistent performance, denoted by a sharp curve in the CDF. The baseline (no scheduling), on the other hand, has a large variance. For comparison, $95^{th}$ percentile of normalized step time in the baseline and TAC are respectively $0.63403$ and $0.99825$. 

\xparagraph{Straggler Effect}: Performance inconsistency creates straggling worker effect when multiple workers have different makespan. As a result, all workers have to wait for the slowest one.  We quantify the straggler time as the maximum time spent by any worker in waiting to the total iteration time (represented in percentage).

In Figure~\ref{fig:compare2} (b), we show the impact of stragglers. Straggler effect is caused by two factors: system-level performance variations and efficiency of scheduling on individual workers. In the baseline, workers follow arbitrary scheduling. Hence, a worker with a bad order forces other workers into a long wait, more than $50\%$ of the total iteration time in some cases. On average, scheduling limits straggler effect with larger benefits in bigger DNNs (higher number of ops).  Enforcing \textit{any} order reduces straggler effect regardless of the quality of the chosen order.

 \section{Conclusion}
In this work, we elucidate the importance of communication scheduling in distributed deep learning systems. We devised a metric for quantifying the efficiency of a given schedule of data transfers and developed two heuristics for efficient scheduling. Through extensive testing of these heuristics across a variety of workloads, we demonstrated that significant gains are achievable through communication scheduling. For a typical DNN training which runs for days to weeks, $20\%$ improvement in iteration throughput can save significant compute power. 

Our study encourages further research in network scheduling for parameter server as well as other unexplored transfer patterns such as all reduce. In future, we can also take into account additional metrics such as congestion from the network fabric for better network performance. The initial results also provide motivation for extending the scheduling to additional resources types such as memory and storage.

\bibliographystyle{sysml2019}
\bibliography{main}

\clearpage
\appendix
\newpage
\section*{Appendix}
\begin{appendix}

\section{DNN Models}
In Table~\ref{tab:models}, we present the model characteristics of $10$ deep learning models used in our evaluation. The number of parameters, total size of all parameters, number of computational operations in inference mode and training mode, and the standard batch size are given below.

\begin{table}[h]
\centering
\begin{tabular}{@{}llllll@{}}
\toprule
\begin{tabular}[c]{@{}l@{}}\smaller{Neural Network}\\ \smaller{Model}\end{tabular} & \begin{tabular}[c]{@{}l@{}}\smaller{\#Par}\\ \smaller{}\end{tabular} & \begin{tabular}[c]{@{}l@{}}\smaller{Total Par}\\ \smaller{Size (MiB)}\end{tabular} & \begin{tabular}[c]{@{}l@{}}\smaller{\#Ops}\\ \smaller{Inference}/\\ \smaller{Training}\end{tabular} & \begin{tabular}[c]{@{}l@{}}\smaller{Batch}\\ \smaller{Size}\end{tabular} \\ \midrule
\smaller AlexNet v2 {\tiny\cite{krizhevsky2014one}}& \smaller 16& \smaller 191.89& \smaller 235/483& ${\scriptstyle 512}$ \\
\smaller Inception v1 {\tiny\cite{inception-v1}}& \smaller 116& \smaller 25.24& \smaller 1114/2246& ${\scriptstyle 128}$ \\
\smaller Inception v2 {\tiny\cite{inception-v2}}& \smaller 141& \smaller 42.64& \smaller 1369/2706& ${\scriptstyle 128}$ \\
\smaller Inception v3 {\tiny\cite{DBLP:journals/corr/SzegedyVISW15}}& \smaller 196& \smaller 103.54& \smaller 1904/3672& ${\scriptstyle 32}$ \\
\smaller ResNet-50 v1 {\tiny\cite{DBLP:journals/corr/HeZRS15}}& \smaller 108& \smaller 97.39& \smaller 1114/2096& ${\scriptstyle 32}$ \\
\smaller ResNet-101 v1 {\tiny\cite{DBLP:journals/corr/HeZRS15}}& \smaller 210& \smaller 169.74& \smaller 2083/3898& ${\scriptstyle 64}$ \\
\smaller ResNet-50 v2 {\tiny\cite{DBLP:journals/corr/HeZR016}}& \smaller 125& \smaller 97.45& \smaller 1423/2813& ${\scriptstyle 64}$ \\
\smaller ResNet-101 v2 {\tiny\cite{DBLP:journals/corr/HeZR016}}& \smaller 244& \smaller 169.86& \smaller 2749/5380& ${\scriptstyle 32}$ \\
\smaller VGG-16 {\tiny\cite{simonyan2014very}}& \smaller 32& \smaller 527.79& \smaller 388/758& ${\scriptstyle 32}$ \\
\smaller VGG-19 {\tiny\cite{simonyan2014very}}& \smaller 38& \smaller 548.05& \smaller 442/857& ${\scriptstyle 32}$ \\ \bottomrule
\end{tabular}
\caption{DNN model characteristics}
\vspace{-5mm}
\label{tab:models}
\end{table}

\section{TIC vs. TAC}

\begin{figure}[b]
    \centering
    \includegraphics[width=\columnwidth]{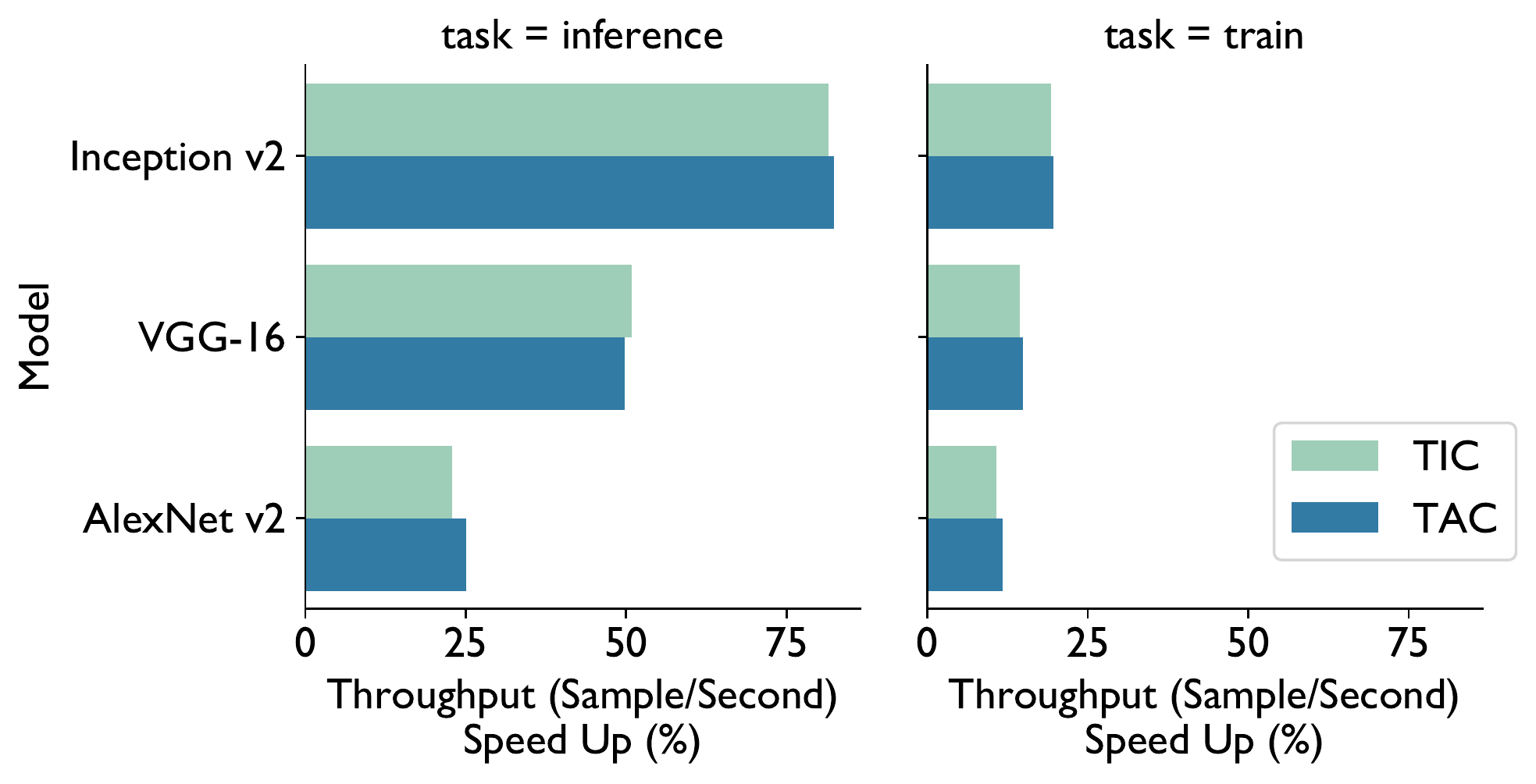}
    \caption{\small Increase in throughput with the scheduling schemes (TIC and TAC) compared to the baseline (no scheduling). Measured on $env_C$ (CPU-Only). }
    \label{fig:cpuRate}
\end{figure}

In Figure~\ref{fig:cpuRate}, we plot the increase in throughput achieved with scheduling in $env_C$ with and without the scheduling schemes (TIC and TAC). We observe that both TIC and TAC offer significant speedup compared to the baseline (no scheduling). Performance of TIC is comparable to that of TAC indicating that we can achieve improved performance without relying on runtime statistics in current models. 

Due to the simplicity of TIC algorithm, we use it as the representative algorithm for scheduling in the cloud GPU environment ($env_G$).

\newpage

\end{appendix} \end{document}